\documentclass[reprint,amsmath,amssymb,aps,prb,superscriptaddress, showpacs]{revtex4-2}
\usepackage{amsmath,amssymb,ascmac,fancybox}
\usepackage{color}
\usepackage{graphicx}
\usepackage{url}
\usepackage{siunitx}
\usepackage{bm}
\usepackage{physics}
\usepackage{mathtools}
\usepackage{hyperref}
\usepackage[T1]{fontenc}
\newcommand{\Ts}{T_s}
\newcommand{\betas}{\beta_s}
\newcommand{\Td}{T_d}
\newcommand{\betad}{\beta_d}
\newcommand{\demonvec}{\vec{\alpha}}
\newcommand{\Ed}{\varepsilon_d}
\newcommand{\demony}{\ket{1}}
\newcommand{\demonn}{\ket{0}}
\newcommand{\Rs}{R_s}
\newcommand{\Rd}{R_d}
\newcommand{\gamR}{\gamma_R}
\newcommand{\gamL}{\gamma_L}
\newcommand{\GamU}{\Gamma^U}
\newcommand{\Gam}{\Gamma^0}
\newcommand{\GamD}{\Gamma_D}
\newcommand{\gamD}[1]{\gamma_{d#1}}

\newcommand{\mgamR}{\hat{\gamma}_R}
\newcommand{\mgamL}{\hat{\gamma}_L}
\newcommand{\mGam}{\hat{\Gamma}}
\newcommand{\mF}{\hat{F}}

\newcommand{\fd}{f_d}
\newcommand{\fdu}{f_d^U}
\newcommand{\fs}{f_s}

\newcommand{\pd}{p_d}
\newcommand{\pdu}{p_d^U}
\newcommand{\vpd}{\vec{p}_d}
\newcommand{\vpdu}{\vec{p}_d^U}

\newcommand{\Efield}{\mathcal{E}}

\newcommand{\mdE}{\Delta \hat{E}}

\newcommand{\Qs}{Q_s}
\newcommand{\Qd}{Q_d}
\newcommand{\dT}{\Delta T}
\newcommand{\dbeta}{\Delta \beta}
\newcommand{\osL}{{\mathcal{L}}} 
\newcommand{\ratioG}{c_{\Gamma}} 

\begin{document}
\title{Feedback-type thermoelectric effect in correlated solids}

\author{Yugo Onishi}
\affiliation{Department of Applied Physics, The University of Tokyo, Hongo, Tokyo, 113-8656, Japan}

\author{Naoto Nagaosa}
\affiliation{Department of Applied Physics, The University of Tokyo, Hongo, Tokyo, 113-8656, Japan}
\affiliation{RIKEN Center for Emergent Matter Science (CEMS), Wako, Saitama, 351-0198, Japan}

\date{\today}

\begin{abstract}
	A new thermoelectric effect mechanism inspired by an autonomous Maxwell's demon [P. Strasberg, G. Schaller, T. Brandes, and M. Esposito, Phys. Rev. Lett. 110, 040601 (2013)] is proposed. In contrast to the former work where a model for microscopic systems is proposed, a specific model for the thermoelectric effect in solid is formulated and its response to the electric field and temperature gradient is calculated in the framework of stochastic thermodynamics. The results show that relatively high $ZT$, which represents efficiency of the thermoelectric material, can be achieved within a range of realistic parameters.
\end{abstract}


\maketitle

\section{Introduction}
The second law of thermodynamics is seemingly violated when feedback is applied to the system at a microscopic level. The existence of such a feedback-giving entity was first proposed by Maxwell and is called Maxwell's demon. Now, this problem has been solved by taking into account the entropy production when the information is erased at the memory, and the second law of thermodynamics has been extended to a more general form ~\cite{Landauer1961, Piechocinska2000, Shizume1995, Maruyama2009, Sagawa2009}. Along with the discussions of Maxwell's demon and its related topics, a formalism called stochastic thermodynamics has been well developed. In stochastic thermodynamics, states including non-equilibrium states are represented by a probability distribution and, heat, work and entropy change are defined at a microscopic level. These microscopic descriptions enable us to investigate microscopic systems including Maxwell's demon and molecular machines, and  general relations such as fluctuation theorems are also established~\cite{Sekimoto2010, Seifert2012}.

Thermal engines using Maxwell's demons have also been proposed. One of the simplest ones is Szilard engine~\cite{Szilard1929,Szilard1964}. This thermal engine converts the information of the system held by Maxwell's demons into work, which has been experimentally realized in previous works~\cite{Serreli2007,Toyabe2010, Koski2014, Koski2015,Vidrighin2016}. In addition, a more autonomous feedback mechanism has been theoretically proposed, for example, by using a single electron transistor to "measure" the state of electrons through interactions and using the results for feedback~\cite{Strasberg2013}. 
A quantum autonomous demon using an exchange interaction between spins~\cite{Ptaszynski2018a} and a demon using non-equilibrium distribution~\cite{Sanchez2019} have also been proposed. We note that all of these examples are microscopic or mesoscopic systems. More general autonomous demon mechanisms have been studied for classical and quantum systems with master equation and quantum master equation\cite{Horowitz2014, Ptaszynski2019}. 

On the other hand, in the field of materials science, materials that exhibit the effect of conversion of heat into electricity in materials, i.e., the thermoelectric effect, have been actively studied. In particular, thermoelectric materials with high efficiency are important from the engineering point of view and are actively explored, 
but the figure of merit, $ZT$, which represents the efficiency, is usually about $ZT \sim 1$ for single-phase thermoelectric materials. $ZT\sim 2.4$ is realized in some tuned nano-structure systems~\cite{Venkatasubramanian2001, Snyder2008, Sootsman2009} and SnSe is recently found to show an unprecedented value of up to $ZT\sim 2.6$~\cite{Zhao2014, Duong2016, Chang2018}.
Theoretically, calculations based on conventional transport theory have been performed taking into account the band structures in most of the cases~\cite{Pei2012, GutierrezMoreno2020}, and the importance for the dimensionality~\cite{Hicks1993, Hicks1993a}, some particular shapes of band structure~\cite{Kuroki2007}, and valley degeneracy~\cite{Pei2011} is discussed. Several formulas have also been derived based on the Kubo formula~\cite{Shastry2009, Peterson2010, Chaikin1976, Shastry2006}. The effects of the electron correlation are often treated by Hubbard model~\cite{Beni1974, Kwak1976, Chaikin1976, Kwak1976a}. 
However, the effects of electron-electron interactions in these calculations have only been incorporated in a limited way, and there has been no discussion of incorporating the effects of feedback as discussed above.

Here we propose that a thermoelectric effect in solid using a Maxwell's demon-like feedback mechanism. A specific model is formulated in the framework of stochastic thermodynamics and the response to the electric field and temperature gradient is calculated. Although the mechanism is similar to the autonomous demon previously proposed~\cite{Strasberg2013}, the present model is for macroscopic systems in contrast to the previous one for a quantum dot. It should be noted that the inversion symmetry and the mirror symmetry are broken in the model, determining the direction of the electric field. The results show that relatively high $ZT$ can be achieved within a range of realistic parameters~\cite{Nagaosa1986b, Painelli1988}.

\section{Model}
\subsection{Model overview}
\begin{figure*}[htbp]
	\centering
	\includegraphics[width=1.6\columnwidth]{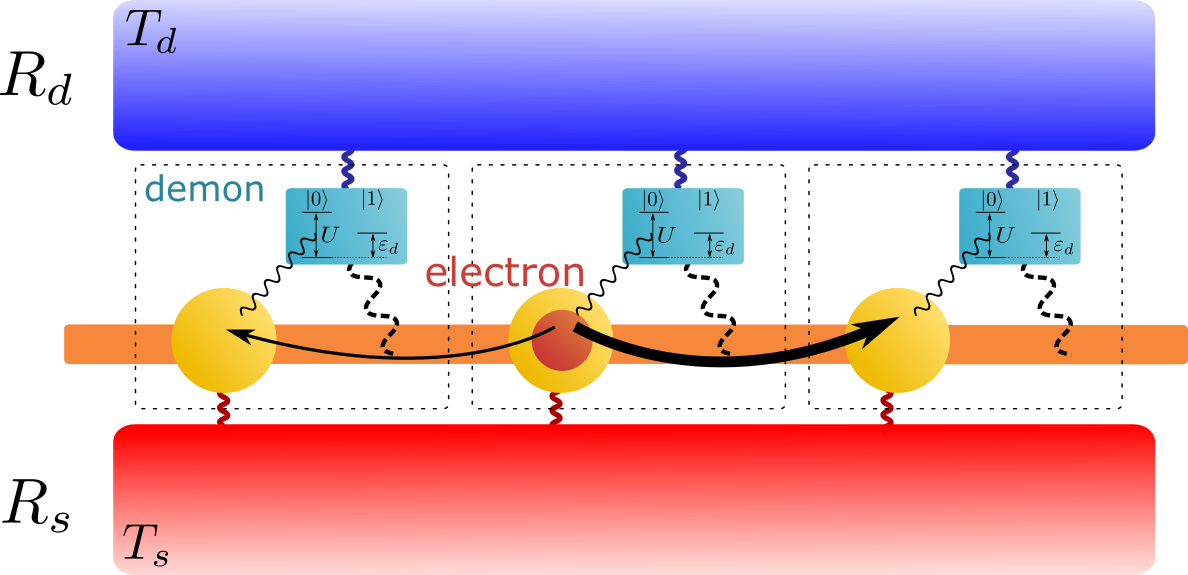}
	\caption{Schematic diagram of the model. The yellow circles represent electron sites and the light blue boxes represent demons. There is one energy level in each electron site, which interacts with a heat bath $\Rs$ with temperature $\Ts$ (inverse temperature $\betas$). For each demon, there are two states, $\demonn$ and $\demony$, which interact with a heat bath $\Rd$ of temperature $\Td$ (inverse temperature $\betad$). The energy level of an electron site and the state $\demonn$ of the demon at the same unit cell interact, and the energy is increased by $U$ when the demon at the site where the electron resides is in $\demonn$.}
	\label{fig:model}
\end{figure*}

Consider a model as shown in Fig.~\ref{fig:model}.
The system consists of $L$ unit cells (the periodic boundary condition), and one unit cell consists of two parts, an electron site and a part we call demon hereinafter. The electron site has a one-electron level in which one electron can occupy (we do not consider spin for simplicity). The demon has a degree of freedom to take one of two states, $\demonn$ and $\demony$. 
Therefore, a state of the system containing $N$ electrons is specified by the position $(x_1, x_2, \dots, x_N)$ of $N$ sites occupied by electrons and the state of $L$ demons $\demonvec = (\alpha_1 , \alpha_2, \dots, \alpha_L)$ (where $\alpha_i = 0,1$). The demon at each site interacts with the electrons if they exist at that site. In particular, we assume that the energy is increased by $U$ because of the interaction only when the demon is in the state $\demonn$. 

The purpose of this paper is to discuss the current flowing among the electron sites when a temperature difference $\Ts-\Td$ is created between the thermal bath felt by demons and the thermal bath felt by electron sites in this model.

To this end, the time evolution of this model is formulated in the framework of stochastic thermodynamics as follows. It is assumed that electrons transit between neighboring electron sites with a certain probability. This transition is supposed to be caused by the thermal bath $\Rs$ at the temperature $\Ts$ with which the electron sites are interacting. 
The transition probability of the electrons will vary with the state of the demon. In particular, the demon at site $i$ controls the probability mainly for the transition from site $i$ to site $i+1$. By this assumption, the system breaks the mirror symmetry, i.e., the right-left symmetry, and feedback in which electrons preferentially move in one direction is realized.
The demon of each unit cell is also supposed to tunnel between the two states $\demonn$ and $\demony$ with a certain probability, and this transition is caused by the thermal bath $\Rd$ at the temperature $\Td$ interacting with the demon. These probabilities satisfy the detailed balance condition except the case of finite electric fields, which guarantees the thermal equilibrium when $\Ts = \Td$.

In order to discuss the thermoelectric effect, it is necessary to calculate the response of the model when an electric field is applied to it. Since our model imposes a periodic boundary condition, it is difficult to explicitly incorporate the electric field as an electrostatic potential into the energy of the system. Instead, the present paper discusses the effects of the electric field by incorporating it into the electron transition probability.

In addition, we use the one-electron approximation below, assuming that the number of electrons $N$ is sufficiently small compared to the number of sites $L$ and that the interactions between electrons are negligible.

In the following we assume that the thermal baths, $\Rs$ and $\Rd$, are realized by fermionic degrees of freedom. However, the extension of the model to the other realization of the baths is trivial.

\subsection{Energy of the system}
When the demon is in state $\demonn$, there is an interaction $U$ between the demon and the electron site. In the range of the one-electron approximation, the energy $E(x, \demonvec)$ when one electron is at position $x$ is given by
\begin{align}
	E(x, \demonvec) &= \sum_{i=1}^L E_i(x, \alpha_i) \\
	E_i(x, \alpha) &= \begin{cases}
		\Ed \delta_{\alpha, 1} & (i\neq x) \\
		\Ed \delta_{\alpha,1} + U\delta_{\alpha, 0} & (i = x) 
	\end{cases}
\end{align}
where $E_i$ is the energy of $i$-th unit cell and $\Ed$ is the energy of demon in state $\demony$. Here, the energy of one electron level of the electron site and the energy of the demon level $\demonn$ are both set to 0.

\subsection{Transition probability}
As mentioned in the model overview, electronic state transitions are triggered by the reservoir $\Rs$, and demon state transitions are triggered by the reservoir $\Rd$. In addition, transitions in which two or more degrees of freedom change at the same time are not considered.

Regarding the transition of the electron, some caution is required in the presence of an electric field. Here, the case where an electric field exists and the case where no electric field exists will be described separately.

\subsubsection{Electron transitions in the absence of an electric field}
An electron can move to a neighboring electron site with a certain probability. This transition probability is determined by the states of the demons at the sites between which the transition takes place. The probability of an electron moving from state $(x, \demonvec)$ to the right (left) $\gamR$ ($\gamL$) is a function of $(\alpha_x, \alpha_{x+1})$ ($(\alpha_x, \alpha_{x-1})$). Here we assume that $\gamR$ and $\gamL$ are given as the following expressions:
\begin{align}
	\gamR(\alpha_x, \alpha_{x+1}) &= \Gamma(\alpha_x) \fs(\Delta E(\alpha_x, \alpha_{x+1})) \\
	\gamL(\alpha_x, \alpha_{x-1}) &= \Gamma(\alpha_{x-1}) \fs(\Delta E(\alpha_x, \alpha_{x-1}))
\end{align}
where $\fs, \Gamma, \Delta E$ are 
\begin{align}
	\fs(\epsilon) &= \frac{1}{1 + e^{\betas\epsilon}} \\
	\Gamma(\alpha_x) &= \begin{cases}
		\GamU & (\alpha_x = 0) \\
		\Gam & (\alpha_x = 1)
	\end{cases} \\
	\Delta E(\alpha_x, \alpha_{x\pm 1}) &= E(x\pm 1, \demonvec) - E(x, \demonvec) \nonumber \\
	& =\begin{cases}
		0 & (\alpha_x = \alpha_{x\pm 1}) \\
		U & (\alpha_x = 1, \alpha_{x\pm 1} = 0) \\
		-U & (\alpha_x = 0, \alpha_{x\pm 1} = 1)
	\end{cases}
\end{align}
and $\GamU$ and $\Gam$ are constants. These two constants characterize the magnitude of the transition probability $\Gamma(\alpha_x)$, which changes depending on the state of the demon. Specifically, the magnitude of the transition probability from $x$ to $x+1$ changes depending on the demon state $\alpha_x$ of the $x$-th unit cell. The fact that the demon affects the magnitude $\Gamma$ only for the transition to its right side, not for the one to its left side, expresses the mirror symmetry breaking of the system and determines the direction of the current.

The difference between $\GamU$ and $\Gam$ also expresses the demon's feedback on electrons. For example, let us consider the case $\GamU < \Gam$. If the demon at a site $x$ is in the state $\demony$, then the probability for an electron at the site $x$ to move to the right is higher than the one when the state of the demon is $\demonn$. 

We use the Fermi-Dirac distribution $\fs$ to define transition probabilities $\gamR, \gamL$ because of the assumption that the reservoir $\Rs$ is realized by fermionic degrees of freedom. The details on this point is given later(Sec.\ref{sec:remark_distribution_function}).

We can also confirm that the transition probability of electrons determined in this way satisfies the condition for detailed balance.
\begin{align}
	\frac{\gamR(\alpha_x, \alpha_{x+1})}{\gamL(\alpha_{x+1},\alpha_x)} = \frac{\fs(\Delta E(\alpha_x, \alpha_{x+1}))}{\fs(-\Delta E(\alpha_x, \alpha_{x+1}))} = e^{-\betas\Delta E(\alpha_x, \alpha_{x+1})} \label{eq:detailed_balance}
\end{align}

\subsubsection{Electron transitions in the presence of an electric field} \label{sec:electron_transition_efield}
Even if there is an electric field, we can treat it exactly the same way as in the previous section if we can express the effect of the electric field as energy. However, in the present model, we cannot define the electrostatic potential well because we impose a periodic boundary condition. Nevertheless we can define the energy change of the system locally and thus we express the electric field by defining $\gamR$, $\gamL$ as follows.
\begin{align}
	\gamR(\alpha_x, \alpha_{x+1}) &= \Gamma(\alpha_x) \fs(\Delta E-v) \\
	\gamL(\alpha_x, \alpha_{x-1}) &= \Gamma(\alpha_{x-1}) \fs(\Delta E+v)
\end{align}
where $v$ is the amount corresponding to the electrostatic potential difference between neighboring sites, expressed as $v = e\Efield a$, where $\Efield$ is the applied electric field, $a$ is the size of the unit cell, and $e$ is the charge of the electron. The electric field is positive when it points to the right direction in Fig.~\ref{fig:model}. 

Note that the detailed balance condition \eqref{eq:detailed_balance} is violated by the introduction of an electric field, since the energy cannot be defined globally.

For convenience of the later calculation, let us rewrite $\gamR, \gamL$ in a matrix form. If we write $\gamR(\alpha_x, \alpha_{x+1})$ in a matrix form with a row index $\alpha_x$ and a column index $\alpha_{x+1}$ and denote it by $\mgamR$, we will get
\begin{align}
	\mgamR &= \mqty(\dmat{\GamU, \Gam})
			\mqty(\fs(-v) & \fs(-U-v) \\
					\fs(U-v) & \fs(-v)) = \mGam \mF(v) \label{eq:mgamR}
\end{align}
where
\begin{align}
	\mGam &= \mqty(\dmat{\GamU, \Gam}) \\
	\mF(v) &= \mqty(\fs(-v) & \fs(-U-v) \\
					\fs(U-v) & \fs(-v))
\end{align}
In the same way, if we display $\gamL(\alpha_x, \alpha_{x-1})$ as a matrix with $\alpha_x$ as the index of the row and $\alpha_{x-1}$ as the index of the column, we get the following.
\begin{align}
	\mgamL &= \mF(-v)\mGam \label{eq:mgamL}
\end{align}

\subsubsection{Transitions of demons}
The transition probability of reversing the state of the demon at site $i$, denoted by $\gamD{i}$, is determined by the location of the reversing demon $i$, its state $\alpha_i$, and the position of the electrons $x$. Here we assume that $\gamD{i}$ has the following form:
\begin{align}
	\gamD{i} &= \gamma(x,i,\alpha_i) = \begin{cases}
		\GamD(1-\fdu) & (x=i, \alpha_i=0)\\
		\GamD\fdu & (x=i, \alpha_i=1) \\
		\GamD\fd & (x\neq i, \alpha_i=0) \\
		\GamD(1-\fd) & (x\neq i, \alpha_i=1)
	\end{cases} \\
	\fd &= \frac{1}{1+e^{\betad \Ed}} \\
	\fdu &= \frac{1}{1+e^{\betad(U-\Ed)}}
\end{align}
The important point is that the equilibrium state of the demon can be changed depending on whether an electron exists at the same site. For example, if $U>\Ed$, the demon prefers the state $\demonn$ in the absence of an electron while the demon tends to be in the state $\demony$ in the presence of an electron at the same site.

Since we are assuming that the reservoirs are realized by fermionic degrees of freedom, the Fermi-Dirac distribution function is used. 
The details on the use of the Fermi-Dirac distribution function is given in Sec.\ref{sec:remark_distribution_function}.

\subsubsection{Remark on the distribution functions in the transition probabilities} \label{sec:remark_distribution_function}
In the model described above, we use the Fermi-Dirac distribution function in the both transition probabilities of the electrons and the demons. The reason is as follows. We assume that both the reservoirs $\Rs$ and $\Rd$ consists of fermionic degrees of freedom. The transition of the electrons or the demons is triggered by the fermions in the reservoirs. However, not all the fermions in the reservoirs can cause transitions. Only the fermions which have enough energy to cause the transitions can make the electrons or the demons change their states. This is why the Fermi-Dirac distribution function appears in the transition probabilities. We also assume that the chemical potential of the reservoirs is equal to 0, i.e., equal to the levels of electron sites. This is because we implicitly assume that the reservoirs are realized by other electrons in the bulk.

\subsection{Master equation}
The time evolution of the system is governed by the master equation determined from the transition probabilities, which are defined in the previous section. In the case of this model, if the probability that the system takes the state specified by the state $(x, \demonvec)$ is written as $p(x, \demonvec)$, the master equation can be written as follows.
\begin{align}
	\pdv{p(x, \demonvec)}{t} &= \gamR(\alpha_{x-1}, \alpha_x)p(x-1, \demonvec) \nonumber \\
	& + \gamL(\alpha_{x+1}, \alpha_x)p(x+1, \demonvec) \nonumber \\
	& -\qty[\gamR(\alpha_x, \alpha_{x+1}) + \gamL(\alpha_x, \alpha_{x-1})]p(x,\demonvec) \nonumber \\
	& + \sum_{i=1}^L \qty[\gamma(x,i,\bar{\alpha}_i)p(x, \demonvec_i')-\gamma(x, i, \alpha_i)p(x,\demonvec)] \label{eq:master_eq}
\end{align}
where $\bar{\alpha}$ denotes the inverted state of $\alpha$. That is, $\bar{\alpha} = 1$ when $\alpha=0$ and $\bar{\alpha} = 0$ when $\alpha=1$. $\demonvec_i'$ in the sum represents the demons' state in which the $i$-th component of $\demonvec$ is reversed, i.e. $\demonvec_i' = (\alpha_1, \dots,\alpha_{i-1}, \bar{\alpha}_i, \alpha_{i+1}, \dots, \alpha_L)$.


\subsection{Demon's feedback in the model}
In the model described above, the demons are expected to give feedback on the electrons' movement. The reason is as follows. 
Let us consider the case of $\GamU \ll \Gam$ and $U\gg \Ed \gg \Td, \Ts$ for simplicity.
Electrons affect the demons' state by the interaction $U$. This interaction makes the demon at the site with an electron prefer the state $\demony$, while the demon is in state $\demonn$ in the absence of electrons.
On the other hand, the states of demons affect the transition probability of electrons. If the demon at a site $x$ is in $\demony$ and the other demons are in $\demonn$, a transition of an electron from a site $x$ to a site $x+1$ is much more likely to occur than a transition from $x$ to $x-1$. Note that the latter transition probability is controlled by the demon at the site $x-1$. As a result, only the demon at the site with an electron becomes state $\demony$ and electrons preferentially move towards the right. This is the feedback effect of the demons we expect. In the next section, we confirm that this expectation is true by calculation using an approximation.

\section{Calculation method and results}
In order to discuss the thermoelectric effect for the model described in the previous section, we want to find the steady state at various temperatures and electric fields. In other words, we want to find the solution of the master equation \eqref{eq:master_eq} when the left side of the master equation is set to zero, but this is generally difficult. Therefore, in the present paper, we assume that the demons respond faster enough than the electrons, i.e., $\GamD \gg\Gam, \GamU$. In this case, we can use an approximation that only the demons relax into equilibrium. Hereinafter, this approximation is called fast demon approximation. At the end of this section, we also investigate $\GamD$ dependence of the linear responses numerically.

\subsection{Fast demon approximation}
Since $\gamma(x, i, \alpha_i) \gg \gamR, \gamL$ when $\GamD \gg \Gam, \GamU$, the master equation on the steady state can be approximated as
\begin{align}
	0 \simeq \sum_{i=1}^L \qty[\gamma(x, i, \bar{\alpha_i})p(x, \demonvec_i') - \gamma(x, i, \alpha_i)p(x, \demonvec)]
\end{align}
The solution of this equation can be easily found by assuming the following form.
\begin{align}
	p(x, \demonvec) &= p(x)\pdu(\alpha_x)\prod_{i\neq x}\pd(\alpha_i)  \label{eq:solution}\\
	\pd(\alpha) &= \begin{cases}
		1-\fd & (\alpha=0) \\
		\fd & (\alpha=1)
	\end{cases}\\
	\pdu(\alpha) &= \begin{cases}
		\fdu & (\alpha = 0) \\
		1-\fdu & (\alpha = 1) 
	\end{cases}
\end{align}
This assumption expresses that the responses of demons are sufficiently fast compared to the response of the electron, so that only the demons relax into equilibrium. 

We can define the effective transition probability to the right and left of the electron, $V_R$ and $V_L$, as follows.
\begin{align}
	V_R &= \sum_{\alpha_x, \alpha_{x+1}} \gamR(\alpha_x, \alpha_{x+1})p(\alpha_x, \alpha_{x+1}|x) \\
	V_L &= \sum_{\alpha_x, \alpha_{x+1}} \gamL(\alpha_x, \alpha_{x-1})p(\alpha_x, \alpha_{x-1}|x)
\end{align}
where $p(\alpha_x, \alpha_{x\pm 1}|x)$ is the conditional probability such that the state of the demons at sites $x,x\pm 1$ are $\alpha_x, \alpha_{x\pm 1}$ under the condition that the electron is at site $x$. This can be expressed using Eq.\eqref{eq:mgamR}, \eqref{eq:mgamL}, \eqref{eq:solution} as, 
\begin{align}
	V_R &= \sum_{\alpha_x, \alpha_{x+1}} \gamR(\alpha_x, \alpha_{x+1})\pdu(\alpha_x)\pd(\alpha_{x+1}) \nonumber \\
	&= (\vpdu)^T \mgamR \vpd = (\vpdu)^T \mGam \mF(v) \vpd \\
	V_L &= (\vpdu)^T \mgamL \vpd = (\vpdu)^T \mF(-v) \mGam \vpd   
\end{align}
where $\vpd, \vpdu$ represents a column vector when the argument of $\pd, \pdu$ is viewed as a subscript.

Using $V_R$ and $V_L$, the master equation for the steady state can be written as
\begin{align}
	0 = \pdv{p(x)}{t} &= V_Rp(x-1) + V_Lp(x+1) - (V_R + V_L)p(x)
\end{align}
Here, $p(x)$ represents the probability such that the electron is at site $x$. The solution of this equation is given by $p(x) = 1/L$. 

In addition, the solution in this fast demon approximation when $\betad = \betas = \betas = \beta$ becomes Boltzmann distribution.  
\begin{align}
	p(x,\demonvec) &\propto e^{\beta U\alpha_x} \prod_{i=1}^L e^{-\beta\Ed\alpha_i} \propto e^{-\beta U(1-\alpha_x)} \prod_{i=1}^L e^{-\beta\Ed\alpha_i} \\
	&= \exp\qty[-\beta\qty(U\delta_{\alpha_x, 0} + \sum_{i=1}^L \Ed\delta_{\alpha_i, 1})] = e^{-\beta E(x,\demonvec)}
\end{align}

\subsection{Calculation of particle current and heat current}
Using $V_R, V_L$ defined in the previous section, the particle flow $I$ can be calculated within the range of the fast demon approximation as follows:
\begin{align}
	I 	&= N(V_R p(x) - V_L p(x+1)) = \frac{N}{L} (V_R- V_L) \nonumber \\
		&= \frac{N}{L}(\vpdu)^T(\mgamR-\mgamL)\vpd \nonumber \\
		&= \frac{N}{L}(\vpdu)^T(\mGam\mF(v)-\mF(-v)\mGam)\vpd
\end{align}
When calculating the heat flow from the heat bath to the system in the fast demon approximation, a little caution is needed: since the solution \eqref{eq:solution} obtained by the fast demon approximation does not contain information on the temporal change in the degrees of freedom of the demon, it is difficult to calculate heat current between the demons and the heat bath $\Rd$ directly in the fast demon approximation.
Therefore, we first calculate the heat current from $\Rs$ to the system, $\Qs$, and thereafter calculate the heat current between $\Rd$ and the system, $\Qd$, using the results of $\Qs$.

In the case without an electric field, the energy the system receives from $\Rs$ when an electron moves from site $x$ to $y=x\pm 1$ is determined by the $\alpha_x$, $\alpha_y$ and given by 
\begin{align}
	\Delta E(\alpha_x, \alpha_y) = \begin{cases}
		0 & (\alpha_x = \alpha_y) \\
		-U & (\alpha_x = 0, \alpha_y = 1)\\
		U & (\alpha_x = 1, \alpha_y = 0)
	\end{cases}
\end{align}
In the presence of an electric field, the energy  received by the system as an electron moves to the right (left), $\Delta E_R$ ($\Delta E_L$), changes by $v$ depending on the direction in which the electron moves.
\begin{align}
	\Delta E_R(\alpha, \beta) &= \Delta E(\alpha, \beta) - v \\
	\Delta E_L(\alpha, \beta) &= \Delta E(\alpha, \beta) + v
\end{align}
Therefore, the heat current $\Qs$ is 
\begin{align}
	\Qs &= N\sum_{x, \demonvec} p(x,\demonvec)\left[\Delta E_R(\alpha_x, \alpha_{x+1})\gamR(\alpha_x, \alpha_{x+1})   \right. \nonumber \\ 
	& \left. + \Delta E_L(\alpha_x, \alpha_{x-1})\gamL(\alpha_x, \alpha_{x-1}) \right] \nonumber \\
	&= N\sum_{\alpha,\beta} \pdu(\alpha)\pd(\beta) \left[\Delta E(\alpha, \beta)\gamR(\alpha,\beta) \right. \nonumber \\ 
	& \left. + \Delta E(\alpha, \beta)\gamL(\alpha,\beta) - v(\gamR(\alpha,\beta)-\gamL(\alpha,\beta)) \right]
\end{align}

To simplify the notation of $\Qs$, we introduce the following operator $*$ between two matrices as follows. Let $A,B$ be the $2\times 2$ matrix. The operator $*$ is defined by:
\begin{align}
	(A*B)_{ij} &= A_{ij} B_{ij}
\end{align}
With this symbol and the matrix representation of $\Delta E, \gamR, \gamL$, the heat current $\Qs$ reads
\begin{align}
	\Qs &= N(\vpdu)^T \qty[\mdE*(\mgamR+\mgamL)] \vpd - Nv((\vpdu)^T(\mgamR-\mgamL) \vpd) \nonumber \\
	&= N(\vpdu)^T \qty[\mdE*(\mgamR+\mgamL)] \vpd - LvI
\end{align}
Next, the heat current $\Qd$ that flows from the system to the heat bath $\Rd$ is calculated. From the energy conservation law, $\Qs$ and $\Qd$ satisfy the following relation.
\begin{align}
	\Qs - \Qd = -LvI
\end{align}
where $-LvI$ on the right-hand side is the amount corresponding to the change in energy per unit time of the system.
Therefore, the heat current flowing into the heat bath $\Rd$ will be
\begin{align}
	\Qd &= \Qs + LvI = N(\vpdu)^T \qty[\mdE*(\mgamR+\mgamL)] \vpd
\end{align}

The above results can be summarized as follows.
\begin{align}
	I &= \frac{N}{L} (\vpdu)^T (\mgamR-\mgamL)\vpd \label{eq:current}\\
	\Qs &= \Qd - LvI \label{eq:Qs}\\
	\Qd &= N(\vpdu)^T \qty[\mdE*(\mgamR+\mgamL)] \vpd \label{eq:Qd}
\end{align}
It can be confirmed that all of these quantities are always zero when $v=0$ and $\Ts = \Td$.

\subsection{Linear response coefficients}
Based on the results of the previous section, we can calculate the linear response coefficients for $\Ts-\Td, v$. The difference $LvI$ between $\Qs$ and $\Qd$ is a nonlinear term and can be ignored here. Thus, we will simply denote both $\Qs$ and $\Qd$ by $Q$ in this section.

In order to obtain the linear response, set $\Td = T, \Ts = T + \dT, \betad = \beta, \betas = \beta + \dbeta$. Also, set $\Delta V = Lv$, $n=N/L$. In this case, $I$ and $Q$ are expressed using Onsager matrix $\osL$ up to the first order of $\dT, v$:
\begin{align}
	\mqty(I\\Q) &= \mqty(L_{ee} & L_{eh} \\
						L_{he} & L_{hh} ) \mqty(Lv/T \\ \dT/T^2) = \osL \mqty(\beta\Delta V \\ -\dbeta)
\end{align}
Each coefficients of $\osL$ can be calculated from Eq.\eqref{eq:current}, \eqref{eq:Qs} and \eqref{eq:Qd}, and we obtain
\begin{align}
	L_{ee} &= \frac{n\Gam}{2L}Y + \frac{nA\Gam}{4L}(1 + \ratioG)X \\
	L_{eh} &= L_{he} = \frac{nU\Gam A}{4}(1-\ratioG) X \\
	L_{hh} &= \frac{NU^2\Gam A}{4}(1 + \ratioG) X
\end{align}
where $A = 1/\cosh^2(\beta U/2), X=\fdu\fd + (1-\fdu)(1-\fd), Y=\frac{\GamU}{\Gam}\fdu(1-\fd)+(1-\fdu)\fd, \ratioG = \GamU/\Gam$.

From this, various physical quantities such as conductivity and thermal conductivity can be obtained as follows.
\begin{align}
	&\text{Conductivity} :& \nonumber \\ 
	&G = \frac{L_{ee}}{T} = \frac{n\Gam}{4LU}\frac{U}{T}(AX(1 + \ratioG) + 2Y) \label{eq:FDA_conductivity}\\
	&\text{Thermal conductivity} :& \nonumber \\ 
	&K = \qty(\frac{Q}{\dT})_{I=0} = \frac{1}{T^2}\qty(L_{hh}-\frac{L_{eh}L_{he}}{L_{ee}}) \nonumber \\ 
	& = \frac{LnU^2\Gam}{2T^2}AX\frac{2AX\ratioG + (1 + \ratioG)Y}{AX(1 + \ratioG)+2Y} \label{eq:FDA_thermal_conductivity}\\
	&\text{Seebeck coefficient} : &\nonumber \\
	&S = -\qty(\frac{\Delta V}{\dT})_{I = 0} = \frac{1}{T}\frac{L_{eh}}{L_{ee}} 
	\nonumber \\ 
	&= \frac{LU}{T}\frac{AX(1-\ratioG)}{AX(1 + \ratioG) + 2Y} \label{eq:FDA_Seebeck}\\
	&\text{Figure of merit} : & \nonumber \\
	&ZT = \frac{GS^2T}{K} = \frac{L_{eh}L_{he}}{L_{ee}L_{hh}-L_{eh}L_{he}} \nonumber \\ 
	&= \frac{AX(1-\ratioG)^2}{4AX\ratioG + 2(1 + \ratioG)Y} \label{eq:FDA_ZT}\\
	&\text{Power factor} : & \nonumber \\
	&PF = GS^2 = \frac{LnU^2\Gam}{4T^3} \frac{A^2 X^2(1-\ratioG)^2}{AX(1+\ratioG) + 2Y} \label{eq:FDA_PF}
\end{align}
We note that the Seebeck coefficient defined here is a off-diagonal component of the Seebeck tensor since the electric current and the heat current are perpendicular to each other.

\subsubsection{Low-temperature limit}
Consider the low temperature limit of $T \ll U, \Ed$. First, in the case of $U>\Ed>0$, $\fd \simeq e^{-\beta\Ed} \to 0$, $\fdu \simeq e^{-\beta(U- \Ed)} \to 0$, $A \simeq 4e^{-\beta U}\to 0$, which leads to $X \to 1$, $Y \simeq \ratioG e^{-\beta(U -\Ed)}\to 0$. Therefore
\begin{align}
	G &\simeq \frac{n\Gam}{2L}\frac{\ratioG e^{-\beta(U-\Ed)}}{T} \to 0 \\
	K &\simeq \frac{LnU^2\Gam}{T^2}(1+\ratioG)e^{-\beta U} \to 0 \\
	S &\simeq \frac{LU}{T}\frac{2(1-\ratioG)}{\ratioG} e^{-\beta\Ed} \to 0 \\
	ZT &\simeq \frac{2(1-\ratioG)^2}{(1+\ratioG)\ratioG}e^{-\beta\Ed} \to 0 \\
	PF &\simeq \frac{2LnU^2\Gam}{T^3} \frac{(1-\ratioG)^2}{\ratioG}e^{-\beta (U+\Ed)} \to 0
\end{align}

On the other hand, if $U<\Ed$, $\fd \sim e^{-\beta\Ed} \to 0$, $1-\fdu \sim e^{-\beta(\Ed-U)} \to 0$, $A \sim 4e^{-\beta U}\to 0$ and thus $X \sim e^{-\beta(\Ed-U)}$, $Y \to \ratioG$.
\begin{align}
	G &\simeq \frac{n\GamU}{2LT} \to \infty\\
	K &\sim \frac{LnU^2\Gam}{T^2}e^{-\beta\Ed}\to 0\\
	S &\sim \frac{2LU}{T} \frac{e^{-\beta\Ed}}{\ratioG} \to 0\\
	ZT &\sim \frac{2(1-\ratioG)^2}{(1+\ratioG)\ratioG}e^{-\beta\Ed} \to 0 \\
	PF &\simeq \frac{2LnU^2\Gam}{T^3} \frac{(1-\ratioG)^2}{\ratioG}e^{-2\beta\Ed} \to 0
\end{align}
The reason why $G\to \infty$ in the limit of $T\to0$ is as follows. In the low temperature limit of $U<\Ed$, the demon's state is effectively fixed at $\demony$, asymptotic to the simple model of electron movement. In this case, electrons cannot be scattered elastically because there is effectively no available state for the scattering. Therefore the lower $T$, the more sensitive the system is to the potential gradient associated with the applied electric field, and the conductivity diverges as $T$ decreases. However, this divergence is considered to be an artifact of the present model since there are other sources of elastic scattering in actual materials, e.g., impurities and defects, which are not incorporated into our idealized model. 

\subsubsection{High-temperature limit}
Consider the high-temperature limit of $T \gg U, \Ed$. In this case, $\fd \simeq 1/2, \fdu \simeq 1/2, A \simeq 1, X \to 1/2, Y \to (\Gam + \GamU)/4\Gam = (1+\ratioG)/4$, and thus
\begin{align}
	G &= \frac{n\Gam}{4LT} \qty(1 + \ratioG) \\
	K &= \frac{Ln\Gam}{4}(\beta U)^2 \frac{\ratioG + \frac{1}{4}(1 + \ratioG)^2}{1 + \ratioG} \\
	S &= L\beta U \frac{1-\ratioG}{1 + \ratioG} \label{eq:Seebeck_HighTlim}\\
	ZT &= \frac{(1 - \ratioG)^2}{(1 + \ratioG)^2 + 4\ratioG} (<1) \\
	PF &= \frac{LnU^2}{16T^3} \Gam \frac{(1-\ratioG)^2}{1+\ratioG}
\end{align}

\subsubsection{In the limit of $\Ed \ll T \ll U$}
Let us consider the case of $\Ed \ll T \ll U$. They lead to $\fd \simeq 1/2, \fdu \simeq e^{-\beta U} \to 0, A \simeq 4e^{-\beta U} \to 0$ and thus $X \to 1, Y \to 1/2$. Therefore, 
\begin{align}
	G &\simeq \frac{n\Gam}{4LT} \\
	K &\simeq \frac{LnU^2\Gam}{T^2} (1+\ratioG)e^{-\beta U} \to 0 \\
	S &\simeq \frac{LU}{T} 4e^{-\beta U}(1-\ratioG) \to 0\\
	ZT &\simeq \frac{4e^{-\beta U}(1-\ratioG)^2}{1 + \ratioG} \to 0 \\
	PF &\simeq \frac{LnU^2\Gam}{T^3} 4e^{-2\beta U}(1-\ratioG)^2 \to 0
\end{align}

\subsubsection{General temperature dependence}
\begin{figure*}[htbp]
	\centering
	\includegraphics[width=2\columnwidth]{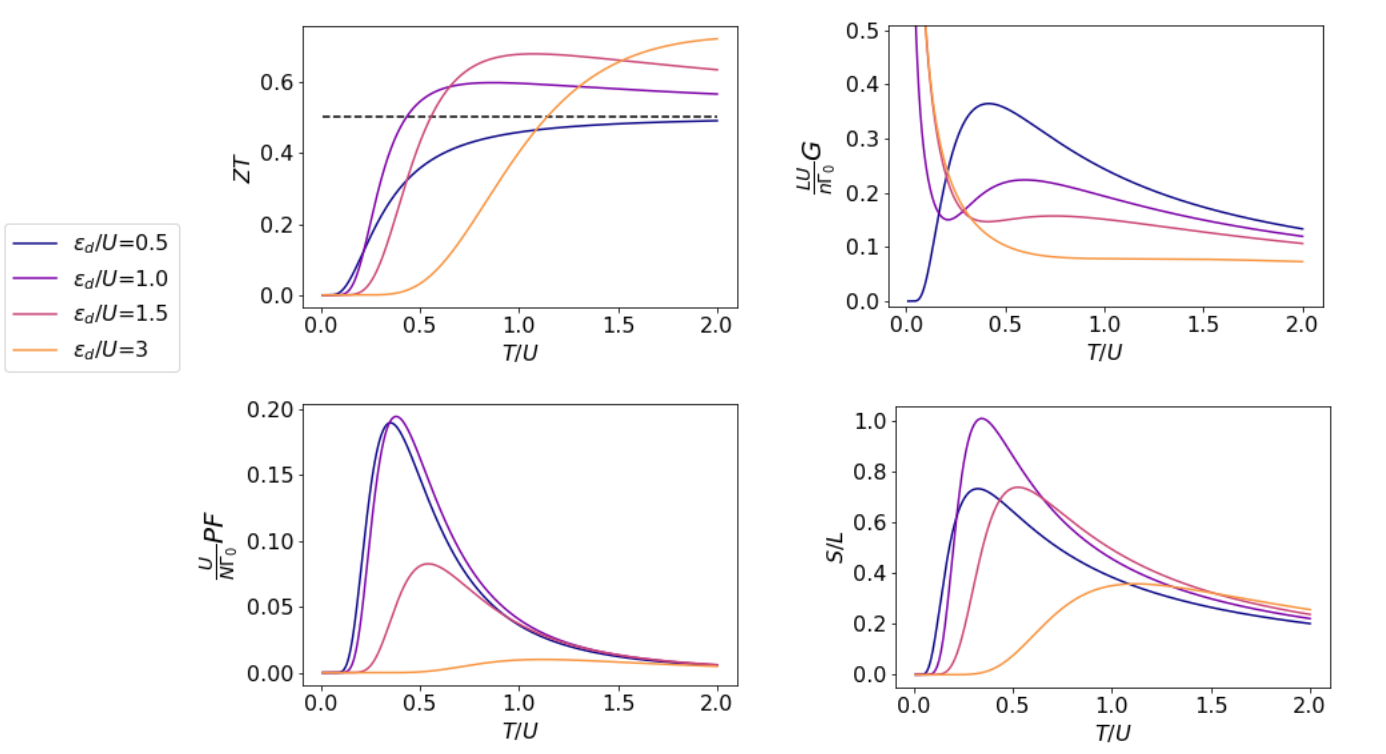}
	\caption{The temperature dependence of $ZT$, normalized power factor $\frac{U}{N\Gam}PF$, normalized conductivity $\frac{LU}{n\Gam }G$, and the normalized Seebeck coefficient $S/L$ when $\ratioG=0.1$ within the fast demon approximation (Eq.\eqref{eq:FDA_conductivity}, \eqref{eq:FDA_Seebeck}, \eqref{eq:FDA_ZT}, \eqref{eq:FDA_PF}). The black dashed line in the $ZT$ plot represents the high-temperature limit of $ZT$.}
	\label{fig:ratio_0.1_ZT_rPF}
\end{figure*}

\begin{figure*}[htbp]
	\centering
	\includegraphics[width=2\columnwidth]{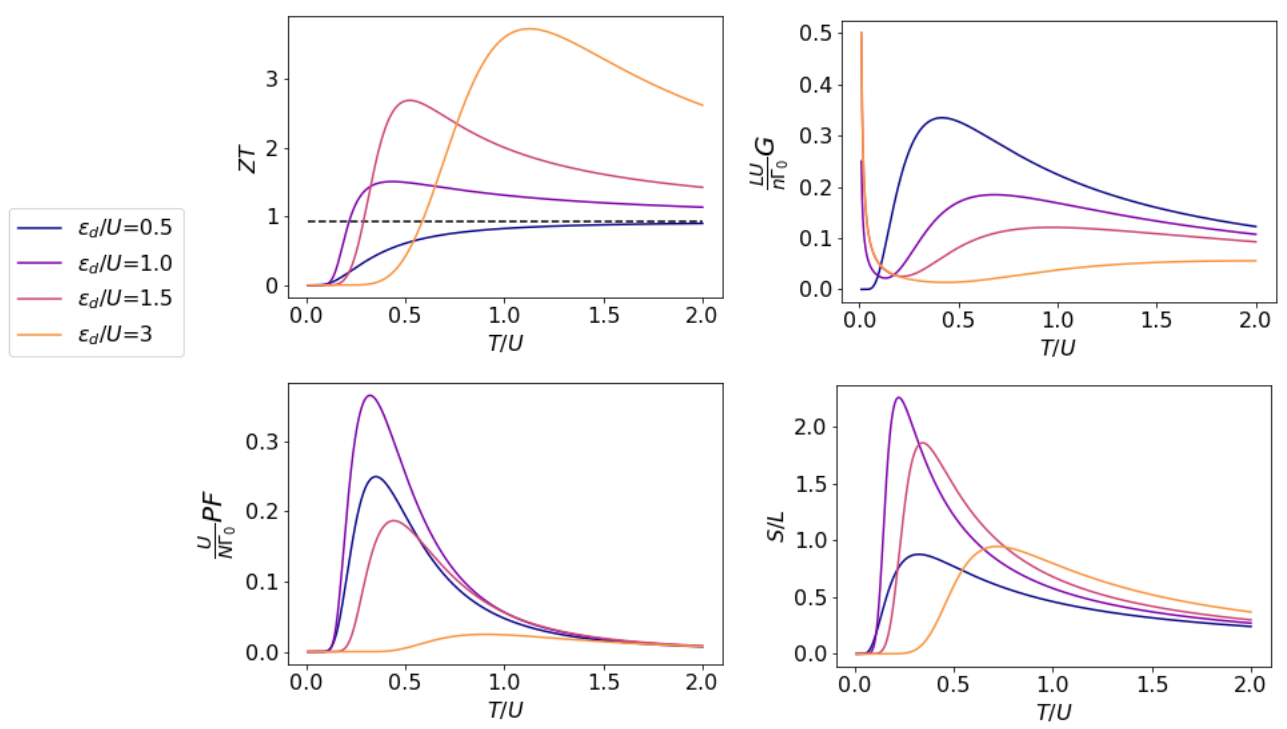}
	\caption{The temperature dependence of $ZT$, normalized power factor $\frac{U}{N\Gam}PF$, normalized conductivity $\frac{LU}{n\Gam }G$, and the normalized Seebeck coefficient $S/L$ when $\ratioG=0.01$ within the fast demon approximation (Eq.\eqref{eq:FDA_conductivity}, \eqref{eq:FDA_Seebeck}, \eqref{eq:FDA_ZT}, \eqref{eq:FDA_PF}). The black dashed line in the $ZT$ plot represents the high-temperature limit of $ZT$.}
	\label{fig:ratio_0.01_ZT_rPF}
\end{figure*}

The general temperature dependence of conductivity, Seebeck coefficient, $ZT$ and $PF$ within the fast demon approximation (Eq.\eqref{eq:FDA_conductivity}, \eqref{eq:FDA_Seebeck}, \eqref{eq:FDA_ZT}, \eqref{eq:FDA_PF}) are plotted in Fig.~\ref{fig:ratio_0.1_ZT_rPF} and Fig.~\ref{fig:ratio_0.01_ZT_rPF}. The smaller the ratio $\ratioG=\GamU/\Gam$, the larger $ZT$ and $PF$. The maximum value of $ZT$ increases as $\Ed/U$ increases, but the maximum value of $PF$ shows a non-monotonic behavior that maximizes around $\Ed/U=1$.

The reason why the maximum value of $ZT$ monotonously increases with respect to $\Ed/U$ even if $\Ed/U$ exceeds 1 is as follows. If $\Ed/U$ is large, the probability that the demon is in state $\demonn$ is high, regardless of the presence of electrons. Therefore, it is difficult to give feedback to the left and right movements of electrons by the demon. However, in rare cases when the demon goes into the state $\demony$, resulting in the movement of electrons, the heat flow into the system from the heat bath at that time becomes very small. By doing so, it is considered that the efficiency of the system as a heat engine becomes very high and the $ZT$ becomes large. Instead, $PF$, which represents the magnitude of output as a heat engine, decreases as $\Ed/U$ increases above a certain level.

The maximum value of $PF$ when $\Ed/U=1$ is considered to be the largest because the Seebeck coefficient is the largest. When $\Ed/U=1$, the effective entropy carried by electrons becomes large, so the Seebeck coefficient may be large.

\subsection{General responses}
Returning to Eq.\eqref{eq:current}, Eq.\eqref{eq:Qs}, and Eq.\eqref{eq:Qd}, we summarize the case of high and low temperature limits and cases where $v$,$\dT$ are too large to be handled by linear response.

\subsubsection{Low-temperature limit}
Here we consider the low-temperature limit, i.e., the case where $\betad \Ed, \betad U, \betas v, \betas U \gg 1$. We assume that $v$ is positive. The behavior changes in each case of $v < U, v > U, \Ed < U, \Ed > U$.
First consider the case of $v<U, \Ed<U$. When $\betas v, \betas U \gg 1$, $\mgamR, \mgamL$ become as follows.
\begin{align}
	\mgamR &= \mqty(\dmat{\GamU, \Gam})
	\mqty(\fs(-v) & \fs(-U-v) \\
			\fs(U-v) & \fs(-v)) \nonumber \\
	&\simeq \mqty(\dmat{\GamU, \Gam})
		\mqty(1 & 1 \\
				0 & 1) 
	= \mqty(\GamU & \GamU \\
			 0	& \Gam) \\
	\mgamL &= \mqty(\fs(v) & \fs(-U+v) \\
					\fs(U+v) & \fs(+v)) 
			  \mqty(\dmat{\GamU, \Gam}) \nonumber \\ 
		&\simeq \mqty(0 & 1 \\
					 0 & 0) \mqty(\dmat{\GamU, \Gam}) 
		= \mqty(0 & \Gam \\
				 0 & 0	)	
\end{align}
and $\vpdu,\vpd$ are
\begin{align}
	\vpd &= \mqty(1-\fd & \fd)^T \simeq \mqty(1 & 0)^T\\
	\vpdu &= \mqty(\fdu & 1-\fdu)^T \simeq \mqty(0 & 1)^T
\end{align}
Therefore, $I, \Qs, \Qd$ are 
\begin{align}
	I &\simeq 0, \quad \Qs \simeq 0, \quad \Qd \simeq 0
\end{align}

Next, we consider the case $v>U, \Ed<U$. In this case $\mgamR, \mgamL$ are  
\begin{align}
	\mgamR &\simeq \mqty(\dmat{\GamU, \Gam})
		\mqty(1 & 1 \\
				1 & 1) 
		= \mqty(\GamU & \GamU \\
				 \Gam  & \Gam) \\
	\mgamL &\simeq \mqty(0 & 0 \\
				  		0 & 0) \mqty(\dmat{\GamU, \Gam})
	 = \mqty(0 & 0 \\
			  0 & 0	) \\
\end{align}
and $\vpd, \vpdu$ are the same as above. The results are
\begin{align}
	I &= \frac{N\Gam}{L} \\
	\Qs &= N\Gam U - N\Gam v \\
	\Qd &= N\Gam U
\end{align}
Therefore, it can be seen that when $\Ed<U$, current suddenly starts to flow when $v$ becomes larger than $U$ in the low temperature limit.

In the case of $\Ed>U$, $\vpdu$ is different, $\vpdu = \mqty(1 & 0)^T$. In this case, for both case $v>U$ and $v<U$, we have
\begin{align}
	I &\simeq \frac{N}{L}\GamU \\
	\Qs &\simeq -Nv\GamU \\
	\Qd &\simeq 0
\end{align}
This means that under $\Ed>U$, the feedback by the demon is no longer working and the currents flow due to the potential gradient or the electric field (which is sufficiently larger than the temperature). Note that the origin of the thermal energy flowing out to the thermal bath $\Rs$ is the work that the external electric field does to the electrons in the system.

\subsubsection{High-temperature limit}
Let us consider the case $\betad \Ed, \betad U, \betas v, \betas U \ll 1$.
Then $\mgamR, \mgamL, \vpd, \vpdu$ are
\begin{align}
	\mgamR &\simeq \frac{1}{2}\mqty(\GamU & \GamU \\ \Gam & \Gam) \\
	\mgamL &\simeq \frac{1}{2}\mqty(\GamU & \Gam \\ \GamU & \Gam) \\
	\vpd &\simeq \vpdu \simeq \mqty(1/2 & 1/2)^T
\end{align}
Therefore, 
\begin{align}
	I &\simeq 0 \\
	\Qs &\simeq 0 \\
	\Qd &\simeq 0 
\end{align}
This corresponds to the fact that the feedback by the demons does not work at high temperatures and the potential gradient due to the electric field is also negligible compared to the temperature.

\subsection{Linear responses beyond the fast demon approximation}
\begin{figure*}[htbp]
	\centering
	\includegraphics[width=2\columnwidth]{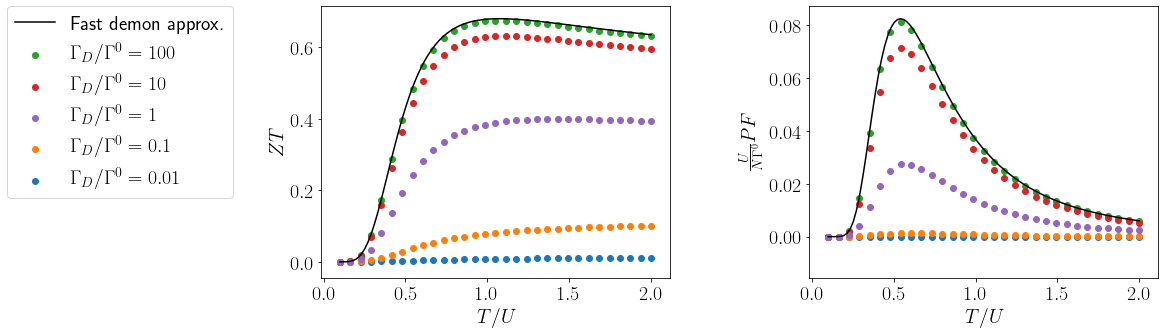}
	\caption{Numerical calculation results of the temperature dependence of $ZT$ and $PF$ when $\ratioG = \GamU/\Gam = 0.1, \Ed/U = 1.5, \Gam = U = 1, L=12$ and different $\GamD/\Gam=0.01,0.1,1,10,100$. The black solid line represents the result of the fast demon approximation (Eq.~\eqref{eq:FDA_ZT}, \eqref{eq:FDA_PF}). The details of the calculation is given in Appendix \ref{Ap:numerical}.}
	\label{fig:ZT_rPF_vs_GamD_GamU_0p1}
\end{figure*}

\begin{figure*}[htbp]
	\centering
	\includegraphics[width=2\columnwidth]{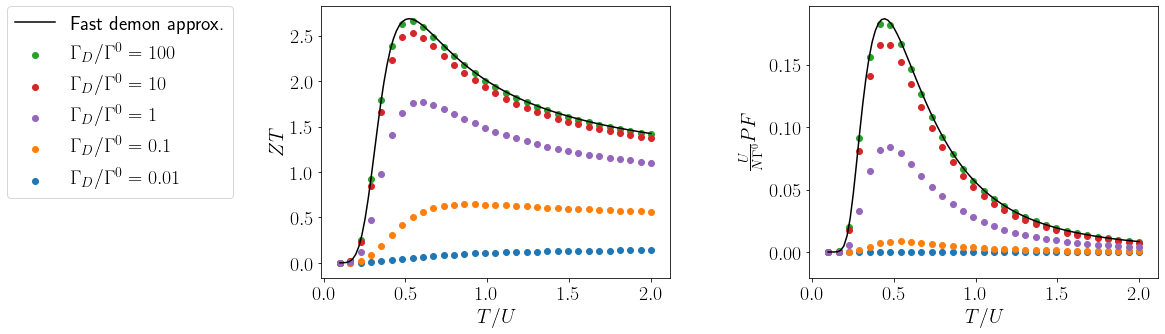}
	\caption{Numerical calculation results of the temperature dependence of $ZT$ and $PF$ when $\ratioG = \GamU/\Gam = 0.01, \Ed/U = 1.5, \Gam = U = 1, L=12$ and different $\GamD/\Gam=0.01,0.1,1,10,100$. The black solid line represents the result of the fast demon approximation (Eq.~\eqref{eq:FDA_ZT}, \eqref{eq:FDA_PF}). The details of the calculation is given in Appendix \ref{Ap:numerical}.}
	\label{fig:ZT_rPF_vs_GamD_GamU_0p01}
\end{figure*}

In this section, we investigate the $\GamD$ dependence of the linear responses.
Here we solve the master equation Eq.\eqref{eq:master_eq} numerically to obtain the probability distribution of the steady state, and calculate the linear responses. For details of the numerical calculation, see Appendix \ref{Ap:numerical}. The results for $L=12, \Ed/U=1.5, \ratioG = \GamU/\Gam=0.01, 0.1$ and various $\GamD$ are shown in Fig.~\ref{fig:ZT_rPF_vs_GamD_GamU_0p1} and Fig.~\ref{fig:ZT_rPF_vs_GamD_GamU_0p01}.

As shown in these results, $ZT$ and $PF$ strongly depend on $\GamD/\Gam$. As $\GamD$ increases, $ZT$ and $PF$ also increase, and when $\GamD/\Gam = 100$ the numerical calculation is almost the same as the result of the fast demon approximation.

This result strongly implies that the time scale separation between demons and electrons is important to realize feedback and improve thermoelectric performances.

\section{Discussion and conclusion}
\subsection{Realization of the model}
First it should be noted that the direction of the heat current and that of the particle current are different from each other. The heat current flows between the reservoirs $\Rs$ and $\Rd$ and thus the heat current is vertical in Fig.~\ref{fig:model}, while the particle current flows between the electron sites, i.e., the particle current is horizontal in Fig.~\ref{fig:model}. 

The model also assumes the system with low conductivity so that the transition of electrons can be treated in stochastic theory.

In order to realize the demons in the present model, the following requirements must be satisfied:~(1) it has two states $\demonn$ and $\demony$, (2) one of them interacts strongly with electrons, (3) the state of the demon affects the transition probability of electrons and gives feedback on the movements of electrons. Any degree of freedom satisfying these requirements can play the role of the demon. Note that the requirement (3) results in the mirror symmetry breaking as in the present model.

For example, suppose that the states $\demonn$ and $\demony$ represent two different one-electron levels. If we assume that the simultaneous occupancy of $\demonn$ and $\demony$ is forbidden because of the strong Coulomb repulsion and that the states interact differently with the electron at the electron site, the above requirements are satisfied. 
Other degrees of freedom, e.g., lattice or electronic spins, may also be used as the demons. Such a situation is expected to be realized in solids if each unit cell has degrees of freedom other than conduction electrons and its symmetry allows, because a change of other degrees of freedom usually affect electrons' dynamics to some extent. Another way to realize the mechanism is to use photons instead of the hot heat reservoir. While the hot reservoir plays the role of injecting energy into the system in our model, the system can get energy from photons with which we can selectively heat only the electrons. In that case, we can make much larger "temperature gradient" in the systems and the responses will be easier to observe in experiment. As for this realization, the model and calculation in the present work should be modified to treat excitation by photons properly, but similar results are expected to be obtained.

It is also important that the response of demons is fast  enough compared to that of electrons. As shown in Fig.~\ref{fig:ZT_rPF_vs_GamD_GamU_0p1} and Fig.~\ref{fig:ZT_rPF_vs_GamD_GamU_0p01}, it is necessary that the time scale of the demons is shorter than, or at least the same as, the time scale of the electrons to realize high $ZT$. 

\subsection{Estimation of $ZT$ and $PF$} \label{subsec:ZT_PF_estimation}
As shown in Fig.~\ref{fig:ratio_0.1_ZT_rPF} and Fig.~\ref{fig:ratio_0.01_ZT_rPF}, $ZT$ strongly depends on the value of $\ratioG=\GamU/\Gam$. If $\ratioG$ is as small as 0.01, $ZT$ can be greater than 2. This is relatively high compared to the known thermoelectric materials~\cite{Wei2020, Snyder2008}.
However, since the current model does not include the effect of heat conduction by the lattice, the actual $ZT$ will be smaller.

Keeping in mind that $PF$ treated up to this point corresponds to the power factor of the entire system, in order to estimate $PF$ specifically, $n=N/L, \Gam/U$, and the lattice constant $a$ (we assume that the system is cubic) have to be estimated.

Here we consider that the demon is realized by electrons in the way described in the previous section, and an organic crystal without inversion symmetry and mirror symmetry is considered as a reference for estimating the value of $n$ and $\Gam/U$. Except for the absence of demon-like degrees of freedom, such systems are expected to satisfy the requirements to apply the model. The order of $\Gam$ is estimated by the hopping parameter $t$. In this case, the order of the power factor per unit length of the system, which we denote here in lower case as $pf$, is estimated to be about $pf \sim \frac{nt}{Ua}$.


As a reference material, we consider TTF-Chloranil(TTF-CA) and other charge-transfer complexes. Note that these compounds break mirror symmetry in two directions and that they also break inversion symmetry by dimerization below a certain transition temperature. These charge-transfer complexes were analyzed with a modified Hubbard model~\cite{Nagaosa1986b, Painelli1988}. In the literature~\cite{Nagaosa1986b}, the transfer integral $t$ and the on-site Coulomb interaction $U$ is estimated to $t\sim\SI{0.2}{eV}, U\sim\SI{1.5}{eV}$, thus $t/U \sim 0.1$. In the literature~\cite{Painelli1988}, $U/t$ is considered to be in the range 10-70. Therefore, we estimate $t/U \sim 0.1$. Furthermore, if we estimate $n/a \sim \SI{1e9}{m^{-1}}$, $pf$ is approximately $\SI{1e8}{m^{-1}} \sim k_B^2/\hbar \times \SI{1e8}{W/mK^2} \sim \SI{100}{\mu W/mK^2}$. In reality, this value is further multiplied by the normalized $PF$ value shown in Fig.~\ref{fig:ratio_0.1_ZT_rPF} and Fig.~\ref{fig:ratio_0.01_ZT_rPF}.

In the above discussion, we use the on-site Coulomb energy for the interaction between demons and electrons, $U$. However, it may be more appropriate to estimate $U$ by nearest neighbor Coulomb interaction. Nonetheless, if we estimate $U$ by nearest neighbor Coulomb interaction in TTF-CA, $U\sim\SI{0.7}{eV}$~\cite{Nagaosa1986b}, the order of magnitude of $pf$ is the same as the above.

\subsection{Comparison with the discussion of the thermoelectric effect using the Hubbard model}
The model assumes a system with low conductivity where the dynamics of electrons can be treated by a stochastic process. In fact, TTF-CA, which is assumed in the estimates above is an insulator. A model often used to discuss this material is a modified Hubbard model. Hubbard model is also used for discussion of Mott insulators, and thermoelectric effects of some organic materials close to Mott transition are discussed with the model~\cite{Beni1974, Kwak1976, Chaikin1976, Kwak1976a}. In the Hubbard model, the Seebeck coefficient is known to be independent of Coulomb repulsion $U$ and temperature $T$ in the high-temperature limit~\cite{Chaikin1976}.


An essential part of the present model is the process of the system's transition to a high-energy state. The high energy state here is such a state that the demon at the electron's site is in the state of $\demonn$, where the energy of the system is higher by $U$. Once the system transitions to this high energy state, the effect of the feedback on the electron transfer due to the relaxation of the state of the demon results in high $ZT$ and $PF$. In order for these effects to be fully reflected in the model, the following three points are important:
\begin{enumerate}
	\item High energy states are achieved with some probability
	\item High energy states can relax by degrees of freedom (playing the role of Maxwell's demon) other than the movement of electrons.
	\item The transition probability of electrons changes depending on the state of the demon
\end{enumerate}

In the Hubbard model, the high energy state changes to the doubly occupied state of the electrons. How to handle this doubly occupied state is important for making the argument in the Hubbard model correspond to the argument in this model.

Regarding these three points, let's see the discussion by Hubbard model in the literatures~\cite{Beni1974, Kwak1976, Chaikin1976}. First, in the discussion by Heikes formula ~\cite{Chaikin1976}, the Seebeck coefficient when the interaction $U$ is sufficiently larger than the temperature $T$ and the temperature is sufficiently larger than the other energy scales, it has been derived that it will be a constant independent of $T, U$. Such discussion does not meet point 1 above and does not reflect the effects of feedback as dealt with in this work.

Also, in other Hubbard model discussions, doubly occupation of electrons is not considered directly. Instead, it is argued that the hopping $t$ is much smaller than the Coulomb interaction between electrons $U$ and is treated by the perturbation expansion by $t/U$. By this perturbation expansion, the transition to the high energy state of the system is considered as an intermediate state in the perturbation. Therefore, it can be said that point 1 is partially satisfied in the sense that it is treated only perturbatively.

However, point 2 and 3 are not satisfied by the usual Hubbard model, where the doubly occupied state is not energetically relaxed by degrees of freedom other than the movement of electrons. Therefore, the essential difference between the Hubbard model and the model in the present paper is that point 2 and 3 above are not satisfied.
On the contrary, when discussing with a model such as the Hubbard model,  the similar result as this model is expected to be obtained if the intermediate state is relaxed by the degree of freedom other than electron movement and feedback is applied.

To conclude, we propose a new mechanism of thermoelectric effect inspired by the concept of Maxwell's demon. A specific model is formulated in the framework of stochastic thermodynamics, and the response to the electric field and temperature gradient is calculated. It is shown that the figure of merit $ZT$ can be relatively high compared to known materials if the demon is realized by electronic degrees of freedom. This mechanism is essentially different from the conventional thermoelectric effects in that it utilizes the correlation effects to realize feedback.

\begin{acknowledgements}
	NN is supported by JST CREST Grant Number JPMJCR1874 and JPMJCR16F1, Japan, and JSPS KAKENHI Grant number 18H03676. 
\end{acknowledgements}

\appendix

\section{Numerical calculation method}\label{Ap:numerical}
To numerically calculate the steady state of the model, we need to solve the following master equation:
\begin{align}
	0 &= \pdv{p(x, \demonvec)}{t} \\ 
	&= \gamR(\alpha_{x-1}, \alpha_x)p(x-1, \demonvec) \nonumber \\
	& + \gamL(\alpha_{x+1}, \alpha_x)p(x+1, \demonvec) \nonumber \\
	& -\qty[\gamR(\alpha_x, \alpha_{x+1}) + \gamL(\alpha_x, \alpha_{x-1})]p(x,\demonvec) \nonumber \\
	& + \sum_{i=1}^L \qty[\gamma(x,i,\bar{\alpha}_i)p(x, \demonvec_i')-\gamma(x, i, \alpha_i)p(x,\demonvec)] \label{eq:steady_state_master_eq} 
\end{align}
This equation can be seen as an eigenvalue problem of $L2^L$ dimension where the eigenvalue is equal to 0. Therefore, we can obtain the steady state by calculating the eigenvector of a certain matrix. However, the dimension of the matrix increases exponentially with the system size $L$, we use translational symmetry to reduce the dimension of the problem from $L2^L$ to $2^L$.

To this end, we assume that the steady state probability distribution $p(x, \demonvec)$ is translationally symmetric, i.e., 
\begin{align}
	p(x+1, \demonvec) &= p(x, T\demonvec)
\end{align}
where $T$ is an operator on demons' state $\demonvec$ defined as
\begin{align}
	T\demonvec &= T(\alpha_1, \alpha_2, \dots, \alpha_L) \label{eq:translational_symmetry} \\
	&= (\alpha_2, \alpha_3, \dots, \alpha_L, \alpha_1) 
\end{align}
Due to the translational symmetry, the steady state probability distribution $p(x, \demonvec)$ satisfies $p(x, \demonvec) = p(0, T^x\demonvec)$. 
Substituting this relation into Eq.~\eqref{eq:steady_state_master_eq}, we obtain
\begin{align}
	0 &= \pdv{p(0, \demonvec)}{t} \\ 
	&= \gamR(\alpha_{-1}, \alpha_0)p(0, T^{-1}\demonvec) \nonumber \\
	& + \gamL(\alpha_{1}, \alpha_x)p(0, T\demonvec) \nonumber \\
	& -\qty[\gamR(\alpha_0, \alpha_{1}) + \gamL(\alpha_0, \alpha_{-1})]p(0,\demonvec) \nonumber \\
	& + \sum_{i=0}^L \qty[\gamma(0,i,\bar{\alpha}_0)p(0, \demonvec_i')-\gamma(0, i, \alpha_i)p(0,\demonvec)] \label{eq:ts_steady_state_master_eq} 
\end{align}
This is an equation for $2^L$ dimensional vector $p(0, \demonvec)$. Combined with Eq.~\eqref{eq:translational_symmetry}, we can calculate the whole steady state probability distribution with Eq.~\eqref{eq:ts_steady_state_master_eq} by solving a $2^L$ dimensional eigenvalue problem. 

\begin{figure*}[htbp]
	\centering
	\includegraphics[width=2\columnwidth]{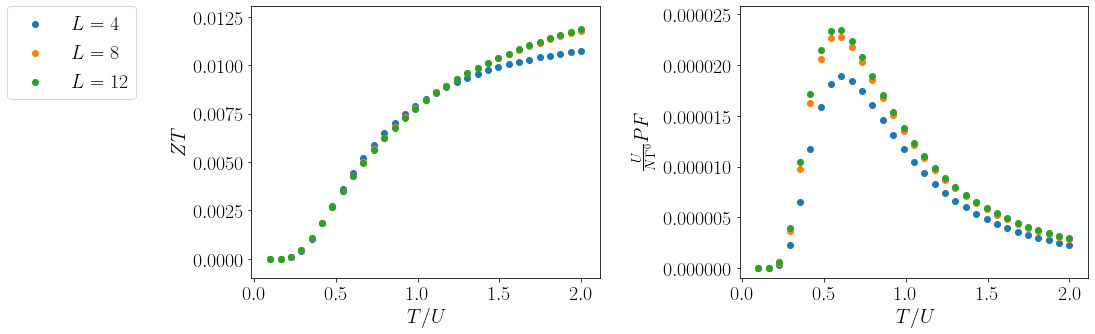}
	\caption{System size ($L$) dependence of the numerical calculations. The parameters used in the calculations are: $\GamU/\Gam=0.1, \GamD/\Gam=0.01, \Ed=1.5, U=\Gam=1$.}
	\label{fig:ZT_PF_vs_L_GamU_0p1}
\end{figure*}

\begin{figure*}[htbp]
	\centering
	\includegraphics[width=2\columnwidth]{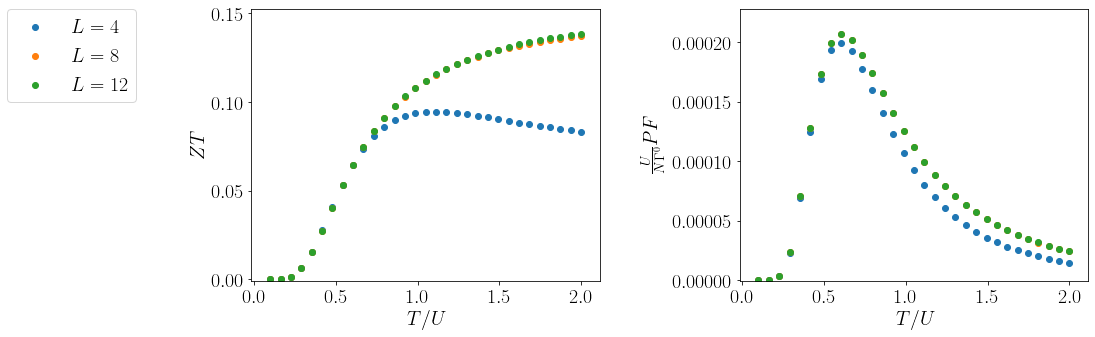}
	\caption{System size ($L$) dependence of the numerical calculations. The parameters used in the calculations are: $\GamU/\Gam=0.01, \GamD/\Gam=0.01, \Ed=1.5, U=\Gam=1$.}
	\label{fig:ZT_PF_vs_L_GamU_0p01}
\end{figure*}

Since we cannot calculate infinite size systems, we must check whether system size $L$ used in calculation is large enough. The results for $\GamD/\Gam = 0.01$ are shown in Fig.~\ref{fig:ZT_PF_vs_L_GamU_0p1} and Fig.~\ref{fig:ZT_PF_vs_L_GamU_0p01}. The figures illustrate that both $ZT$ and $PF$ almost converge at $L=12$. Therefore, we used $L=12$ in the calculations for Fig.~\ref{fig:ZT_rPF_vs_GamD_GamU_0p1} and Fig.~\ref{fig:ZT_rPF_vs_GamD_GamU_0p01}.

\bibliography{references_2nd_submit}

\end{document}